\documentclass[runningheads]{llncs}
\usepackage[T1]{fontenc}
\usepackage{graphicx}
\usepackage{array}
\usepackage{amssymb}
\usepackage{amsmath}
\usepackage{amsfonts}
\usepackage{amssymb}
\usepackage{multirow}
\usepackage{multicol}
\usepackage{mathrsfs}
\usepackage{hyperref}
\hypersetup{
    colorlinks=true,
    citecolor=blue,
    linkcolor=red,
    filecolor=magenta,      
    urlcolor=cyan,
    }
    
\begin{document}
\title{Active Source-free Domain Adaptation in Open-set Medical Image Segmentation via Decomposed Uncertainty and Prototype Discrepancy}
\titlerunning{ASFOSDA for Medical Image Segmentation}
%
%
%


\author{Jin Yang\inst{1} \and Yichi Zhang \inst{2} \and Peijie Qiu\inst{3} \and Xiaobing Yu\inst{3}} 

\authorrunning{J. Yang et al.}
%
\institute{Department of Radiation Oncology, Icahn School of Medicine at Mount Sinai, New York, NY, USA, 10029 \and
School of Data Science, Fudan University, Shanghai, China \and
Mallinckrodt Institute of Radiology, Washington University School of Medicine in St. Louis, St. Louis, MO, USA, 63110 \\
\email{yang.jin@wustl.edu}}
  
\maketitle              
\begin{abstract}
Deep learning (DL) methods are challenged to demonstrate robust performance across different segmentation datasets due to domain shifts, but active domain adaptation techniques enhance their generalization performance by querying a few samples from target domains for adaptation training. However in clinical practice, target domains often include private classes of new anatomical structures or pathologies that are not presented in the source data, and existing methods implement closed-set segmentation where source and target domains have the same segmentation classes. Additionally, source data are often inaccessible during adaptation due to strict data privacy regulations. To address these limitations, we propose an Active Source-free Open-set Domain Adaptation (ASFOSDA) method which is the first work to implement active learning for adapting DL models in open-set medical image segmentation without the access to source data. This method employs an active open-set query strategy to select the most informative target samples for training models based on Class-aware Decomposed Uncertainty (CDU) and Class-agnostic Prototype Discrepancy (CPD). CDU measures sample aleatoric uncertainty and model epistemic uncertainty by employing test time augmentation in stochastic processes. CPD measures cross-domain and self-domain discrepancy for selecting diverse samples. Subsequently, to boost the adaptation performance by enhancing training samples, a Target-refined Self-training strategy is proposed to generate high-quality pseudo labels for unselected samples, thus combining them with labeled samples for a semi-supervised training. We evaluated our method on cross-domain open-set volumetric medical image segmentation tasks, and it outperformed state-of-the-art adaptation methods.


\end{abstract}
\section{Introduction}
Deep learning (DL) methods achieve great success in automatic organ segmentation, but they are limited from generalizing across different data sources due to domain shifts \cite{guan2021domain,yang2025adapting}. Domain adaptation (DA) methods enhance DL methods’ generalizability to address these shifts by transferring knowledge from labeled source domains to unlabeled target domains \cite{zhang2025enhancing}. However, most DA methods implement "closed set" segmentation \cite{wang2019abdominal,yang2026d}. In clinical practice, target domains often include private or new classes of anatomical structures or pathologies, and these classes are not delineated in the source data. Thus, existing DA methods may lead to negative transfer when aligning features of unknown classes from source to target domains, degrading model adaptation performance \cite{wang2021learning}. Additionally, strict data privacy regulations may lead to the inaccessibility of source data. Source-free domain adaptation (SFDA) problems are those of adapting models to target domains without the access to the source data \cite{bateson2022source}, and active learning (AL) is their feasible solution \cite{yang2017suggestive,gaillochet2023active,zhou2024sbc}. AL employs a query strategy to select informative samples for annotation from target domains, thus training models to maximize adaptation performance with minimal annotation overhead \cite{wang2024dual,luo2024uncertainty,yang2025continual,yang2025active}. However, these active source-free domain adaptation (ASFDA) methods lack mechanisms to evaluate the information level of new classes, and no work is implemented to adapt models across domains in open-set medical image segmentation.

To address these limitations, we propose a novel \textbf{Active Source-free Open-set Domain Adaptation} (ASFOSDA) method for medical image segmentation. It employs a novel \textbf{active open-set query strategy} to select informative samples for model adaptation by evaluating the information levels of known and unknown classes based on two metrics: \textbf{Class-aware Decomposed Uncertainty} (CDU) and \textbf{Class-agnostic Prototype Discrepancy} (CPD). First, CDU measures class-aware aleatoric uncertainty (CAU) and class-aware epistemic uncertainty (CEU) from stochastic processes by employing test time augmentation within the target domain. CAU calculates predictive entropy from probability maps of target classes that are unknown to source data in the original and augmented images. Thus, querying samples by CAU to train models improves their reliability on modeling target patterns. CEU calculates variances of predictive energy for unknown classes between the original and augmented images, and querying samples by CEU for training models improves their capabilities of generalizing across new domains. Second, CPD selects representative samples from high-uncertainty ones to train models for further learning target data manifold. CPD measures source-target domain discrepancy by evaluating semantic distance between low-uncertainty and high-uncertainty samples; to avoid selecting samples with redundant knowledge, it also measures semantic dissimilarity among high-uncertainty target samples. Additionally, to improve adaptation performance by enhancing training samples, we implement a \textbf{Target-refined Self-training} to generate high-quality pseudo labels for unselected samples, thus adapting models with labeled samples and pseudo-labeled samples via a semi-supervised training. We evaluated the effectiveness of ASFOSDA on 3D volumetric multi-organ segmentation in CT and MR images by adapting models to segment organs with known and unknown classes. Our method achieved superior performance than state-of-the-art (SOTA) SFDA and ASFDA methods.

\begin{figure*}[t]
    \centering
    \includegraphics[width=\linewidth]{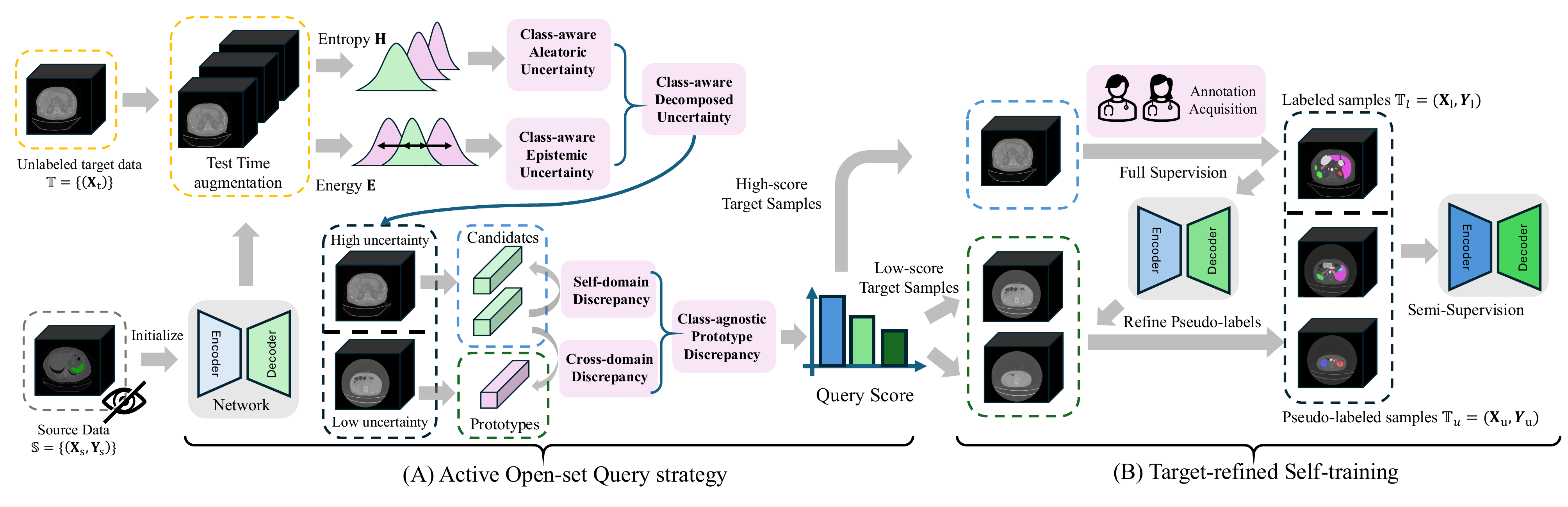} \\
    \caption{\textbf{Active Source-free Open-set Domain Adaptation in medical image segmentation}. \textbf{(A) Active Open-set Query strategy:} After test time augmentation was employed, Class-aware Aleatoric Uncertainty and Class-aware Epistemic Uncertainty were measured, and combined to measure Class-aware Decomposed Uncertainty. Class-agnostic Prototype Discrepancy measured self-domain discrepancy and cross-domain discrepancy among low- and high-uncertainty samples. \textbf{(B) Target-refined Self-training:} Target samples with high scores were selected for annotation $\mathbb{T}_l=\{(\boldsymbol{X}_l,\boldsymbol{Y}_l)\}$ and training the network, and then this network regenerated and refined pseudo-labels for unselected samples with low scores $\mathbb{T}_u=\{(\boldsymbol{X}_u,\boldsymbol{Y}_u)\}$. Finally, labeled and pseudo-labeled samples were combined for semi-supervised training.}
    \label{fig1}
\end{figure*}

\section{Methods}
A source domain $\mathbb{S}$ includes $\mathcal{N}_s$ images $\boldsymbol{X}_s=\{x_s|1\leq s\leq \mathcal{N}_s\}$ with annotation $\boldsymbol{Y}_s=\{y_s|1\leq s\leq \mathcal{N}_s\}$ of $\mathcal{C}_s$ organs, and a target domain $\mathbb{T}$ includes $\mathcal{N}_t$ images $\boldsymbol{X}_t=\{x_t|1\leq t\leq \mathcal{N}_t\}$ without annotation. The goal of ASFOSDA is to derive a segmentation model $\mathscr{F}(\boldsymbol{\Theta})$ with source prior knowledge of segmenting $\mathcal{C}_s$ organs to generate predicted segmentation masks $\boldsymbol{Y}_t=\{y_t|1\leq t\leq \mathcal{N}_t\}$ of $\mathcal{C}_t$ organs in target images $(\mathcal{C}_t>\mathcal{C}_s)$, thus minimizing the risk on the target domain without access to the source domain (Fig.~\ref{fig1}). Before AL query, the model $\mathscr{F}(\boldsymbol{\Theta})$ is initialized by source data $\mathbb{S}=\{(\boldsymbol{X}_s,\boldsymbol{Y}_s)\}$. A labeled target set is empty as $\mathbb{T}_l=\varnothing$, and an unlabeled target set is initialized on the whole target set as $\mathbb{T}_u=\mathbb{T}=\{\boldsymbol{X}_t\}$. Subsequently, with a pre-defined query budget $\mathcal{N}_l$ $(\mathcal{N}_l<<\mathcal{N}_t)$, ASFOSDA employs an active open-set query strategy to select $\mathcal{N}_l$ images $\boldsymbol{X}_l=\{x_l|1\leq l\leq\mathcal{N}_l\}$ for annotation $\boldsymbol{Y}_l=\{y_l|1\leq l\leq\mathcal{N}_l\}$ into the labeled target set $\mathbb{T}_l$ from the unlabeled target set $\mathbb{T}_u$, where $\mathbb{T}_l=\mathbb{T}_l\cup\{(\boldsymbol{X}_l,\boldsymbol{Y}_l)\}$ and $\mathbb{T}_u=\mathbb{T}_u\backslash\{\boldsymbol{X}_l\}=\{\boldsymbol{X}_u\}$. Subsequently, a Target-refined Self-training strategy adapts the model by selected samples with their annotation $\{(\boldsymbol{X}_l,\boldsymbol{Y}_l)\}$ and unselected samples with their refined pseudo labels $\{(\boldsymbol{X}_u,\boldsymbol{Y}_u)\}$ in a semi-supervised training.

\subsection{Active open-set query strategy}
\subsubsection{Class-aware Decomposed Uncertainty.} CDU evaluates predictive uncertainty by decomposing it into the measurement of class-aware aleatoric uncertainty (CAU) and class-aware epistemic uncertainty (CEU). CAU and CEU are measured from stochastic processes by employing test time augmentation (TTA). For an unlabeled target sample $x_u$ from the set $\boldsymbol{X}_u$, TTA generates its augmented variations $\boldsymbol{x}_u^M=\{x_u^1,...,x_u^m\}$ by applying image augmentation techniques, including flipping, rotations, brightness shifts, contrast shifts, and Gaussian noising.

The source-trained segmentation model $\mathscr{F}(\boldsymbol{\Theta})$ is utilized to generate probability maps of the original sample as $\boldsymbol{P}(x_u;\boldsymbol{\Theta})=\{p_1,...,p_C\}\in\mathbb{R}^{C\times H\times W\times D}$ for all target segmentation classes $\mathcal{C}_t$, where $H, W, D$ represent height, width, depth of the image, respectively. The number of channels $C$ is equal to the number of segmented organs $\mathcal{C}_t$ ($C=\mathcal{C}_t$), and the target domain includes known source classes $\mathcal{C}_s$ and unknown target classes $\mathcal{C}_u=\mathcal{C}_t\backslash\mathcal{C}_s=\{c_1^u,...,c_{t-s}^u\}$. Class-aware predictive entropy $\textrm{H}(x_u)$ of the original sample $x_u$ is calculated for unknown classes $\mathcal{C}_u$ from corresponding probability maps $\boldsymbol{P}_u(x_u;\boldsymbol{\Theta})=\{p_1^u,...,p_{t-s}^u\}$ as
\begin{align}
    \textrm{H}(x_u) =-\boldsymbol{P}_u(x_u;\boldsymbol{\Theta}) \log \boldsymbol{P}_u(x_u;\boldsymbol{\Theta})=-\sum_{i=1}^{t-s}p_i^u\log p_i^u.
\end{align}
Subsequently, the model generates probability maps $\boldsymbol{P}(x_u^m;\boldsymbol{\Theta})\in\mathbb{R}^{C\times H\times W\times D}$ for augmented variations $\boldsymbol{x}_u^M=\{x_u^1,...,x_u^m\}$, and their class-aware predictive entropy $\textrm{H}(\boldsymbol{x}_u^M)$ is calculated for unknown classes $\mathcal{C}_u$ from their corresponding probability maps $\{\boldsymbol{P}_1(x_u^1;\boldsymbol{\Theta}),...,\boldsymbol{P}_m(x_u^m;\boldsymbol{\Theta})\}=\{\{p_1^1,...,p_{t-s}^1\},...,\{p_1^m,...,p_{t-s}^m\}\}$ as
\begin{align}
    \textrm{H}(\boldsymbol{x}_u^M)&=\sum_{j=1}^m\textrm{H}(x_u^j) =-\sum_{j=1}^m\boldsymbol{P}_j(x_u^j;\boldsymbol{\Theta}) \log \boldsymbol{P}_j(x_u^j;\boldsymbol{\Theta})=-\sum_{j=1}^m\sum_{i=1}^{t-s}p_i^j\log p_i^j.
\end{align}
Finally, CAU is calculated by averaging the predictive entropy of the original sample $x_u$ and its augmented variations $\boldsymbol{x}_u^M=\{x_u^1,...,x_u^m\}$ as
\begin{align}
    \textrm{CAU}(x_u)= \frac{1}{1+m}\big[\textrm{H}(x_u)+\textrm{H}(\boldsymbol{x}_u^M)\big].
\end{align}
Energy measures how well models align with the domain data, thus evaluating epistemic uncertainty \cite{xie2022active}. The model $\mathscr{F}(\boldsymbol{\Theta})$ generates predicted logits of the original sample $\boldsymbol{Z}(x_u;\boldsymbol{\Theta})=\{z_1,...,z_C\}\in\mathbb{R}^{C\times H\times W\times D}$ for all target segmentation classes $\mathcal{C}_t$. Class-aware predictive energy $\textrm{E}(x_u)$ of the original sample $x_u$ is calculated for unknown classes $\mathcal{C}_u$ from logits $\boldsymbol{Z}_u(x_u;\boldsymbol{\Theta})=\{z_1^u,...,z_{t-s}^u\}$ as
\begin{align}
    \textrm{E}(x_u)=-T\cdot \log e^{\boldsymbol{Z}_u(x_u;\boldsymbol{\Theta})/T}=-T\cdot \log\sum_{i=1}^{t-s}e^{z_i^u/T}.
\end{align}
Subsequently, the model generates predicted logits $\boldsymbol{Z}(x_u^m;\boldsymbol{\Theta})\in\mathbb{R}^{C\times H\times W\times D}$ for augmented variations $\boldsymbol{x}_u^M=\{x_u^1,...,x_u^m\}$, and their class-aware predictive energy $\textrm{E}(\boldsymbol{x}_u^M)$ is calculated for unknown classes $\mathcal{C}_t^u$ from their corresponding logit maps $\{\boldsymbol{Z}_1(x_u^1;\boldsymbol{\Theta}),...,\boldsymbol{Z}_m(x_u^m;\boldsymbol{\Theta})\}=\{\{z_1^1,...,z_{t-s}^1\},...,\{z_1^m,...,z_{t-s}^m\}\}$ as
\begin{align}
    \textrm{E}(\boldsymbol{x}_u^M)=\sum_{j=1}^m\textrm{E}(x_u^j)=-\sum_{j=1}^mT\cdot \log e^{\boldsymbol{Z}_j(x_u^j;\boldsymbol{\Theta})/T}=-\sum_{j=1}^mT\cdot \log\sum_{i=1}^{t-s}e^{z_i^j/T}.
\end{align}
$T$ is a temperature parameter, and set to 1 as commonly used. The average energy among the original sample and its augmented variations is calculated as $    \Bar{\textrm{E}}=\frac{1}{1+m}\big(\textrm{E}(x_u)+\textrm{E}(\boldsymbol{x}_u^M)\big)$. Thus, CEU is calculated by calculating the variances of predictive energy between the original sample $x_u$ and its augmented variations $\boldsymbol{x}_u^M=\{x_u^1,...,x_u^m\}$ as
\begin{align}
    \textrm{CEU}(x_u)=\frac{1}{1+m}\bigg[\big(\textrm{E}(x_u)-\Bar{\textrm{E}}\big)^2+\sum^m_{j=1}\big(\textrm{E}(x_u^j)-\Bar{\textrm{E}}\big)^2\bigg].
\end{align}
Higher CAU and CEU scores demonstrate higher aleatoric and epistemic uncertainty, respectively. When combining them to calculate the CDU score, we apply a min-max normalization and a uniform Quantile transformation to ensure they are in the same range as
\begin{align}
    \textrm{CDU}(x_u)=Norm(\textrm{CAU}(x_u))+Norm(\textrm{CEU}(x_u)).
\end{align}

\subsubsection{Class-agnostic Prototype Discrepancy.}
A source-trained model demonstrates low predictive uncertainty on target samples if it has been equipped with sufficient knowledge to extract accurate patterns from these samples. Thus, these samples are identified as source-like samples, and their feature embeddings generated by the source-trained model represent source domain knowledge. Thus, the encoder of the segmentation model $\mathscr{F}(\boldsymbol{\Theta})$ is employed to generate feature embeddings of $K$ target samples with low CDU score as prototypes $\{\boldsymbol{L}_1,...,\boldsymbol{L}_K\}$ to represent source knowledge. In contrast, target samples with high uncertainty exhibit a significant distribution shift from the source domain, and these target samples are identified as target-specific samples to represent target knowledge. Thus, querying these samples for training enables models to learn target-specific knowledge for adaptation. The encoder generates feature embeddings of $K$ target samples with high CDU score as target query candidates $\{\boldsymbol{H}_1,...,\boldsymbol{H}_K\}$. Subsequently, to select samples for describing knowledge gaps between source and target domains, the cross-domain discrepancy (CDD) is measured for high CDU sample $x_u$ by calculating the semantic distance between its candidate $\boldsymbol{H}_u$ and all prototypes $\{\boldsymbol{L}_1,...,\boldsymbol{L}_K\}$ as
\begin{align}
    \textrm{CDD}(x_u) = \frac{1}{K}\sum_{j=1}^{K}\big(1-\frac{\boldsymbol{H}_u\cdot\boldsymbol{L}_j}{||\boldsymbol{H}_u||\cdot||\boldsymbol{L}_j||}\big).
\end{align}
To avoid selecting samples with redundant knowledge within target domain, the self-domain discrepancy (SDD) for the sample $x_u$ is measured by calculating the semantic dissimilarity between its candidate $\boldsymbol{H}_u$ and other candidates as
\begin{align}
    \textrm{SDD}(x_u) = \frac{1}{K-1}\sum_{i=1}^{K-1}\big(1-\frac{\boldsymbol{H}_u\cdot\boldsymbol{H}_i}{||\boldsymbol{H}_u||\cdot||\boldsymbol{H}_i||}\big).
\end{align}
The CPD score of the sample $x_u$ is calculated by combining CDD and SDD as
\begin{align}
    \textrm{CPD}(x_u) = \textrm{CDD}(x_u)+\textrm{SDD}(x_u).
\end{align}

\subsection{Target-refined Self-training}
To improve adaptation performance by enhancing training samples, the source-trained model is adapted by queried samples with their annotation and unselected samples with their pseudo labels via a semi-supervised training rather than solely using queried samples. However, pseudo labels generated by the source-trained model are in low-quality, and using these pseudo-labeled samples degrades adaptation performance. To address these limitations, the Target-refined self-training strategy implements semi-supervised adaptation in two stages. First, it trains the segmentation model $\mathscr{F}(\boldsymbol{\Theta})$ with selected target samples with their annotation $\{\boldsymbol{X}_l,\boldsymbol{Y}_l\}$ on the labeled target set $\mathbb{T}_l$ via full supervision with a supervised loss function $\mathcal{L}_{sup}(\boldsymbol{Y}_l,\boldsymbol{P}_l)$. Subsequently, the target-trained model is employed to regenerate and refine pseudo-labels for unlabeled target samples $\boldsymbol{X}_u$ on the unlabeled target set $\mathbb{T}_u$. These pseudo-labeled images $\{\boldsymbol{X}_u,\boldsymbol{Y}_u\}$ and labeled images $\{\boldsymbol{X}_l,\boldsymbol{Y}_l\}$ are combined to further adapt the model in a semi-supervised training via a joint loss function $\mathcal{L}=\mathcal{L}_{sup}(\boldsymbol{Y}_l,\boldsymbol{P}_l)+\mathcal{L}_{unsup}(\boldsymbol{Y}_u,\boldsymbol{P}_u)$.

\section{Experiments}
\subsubsection{Datasets.} (i) \textbf{MSD Spleen}: 41 CT scans with spleen annotations \cite{antonelli2022medical}. (ii) \textbf{FLARE}: 361 CT scans with annotations of four abdominal organs (spleen, liver, kidneys, and pancreas) \cite{ma2022fast}. (iii) \textbf{AMOS CT}: 300 CT scans with annotations of 14 organs (spleen, liver, kidneys, pancreas, gall bladder, esophagus, stomach, aorta, postcava, right adrenal gland, left adrenal gland, duodenum, bladder, and prostate/uterus), where right and left kidneys were merged to maintain consistency with FLARE. (iv) \textbf{Abdominal MR}: 30 MR images with annotations of eight anatomical structures, including spleen, right kidney, left kidney, liver, pancreas, gallbladder, aorta, and inferior vena cava \cite{zhou2025mrannotator}. (v) \textbf{AMOS MRI}: 60 MR scans with 13 organ categories where bladder and prostate/uterus were excluded due to extremely imbalanced annotation \cite{ji2022amos}.

\subsubsection{Implementation details.} The experiments were implemented using PyTorch on NVIDIA Tesla A100 PCI-E Passive Single GPU. The 3D U-Net and Swin UNETR were used as segmentation networks \cite{ronneberger2015u,hatamizadeh2021swin}. The combination of dice loss and cross entropy loss was used in supervised training and self-training. Networks were initialized on source data for 1000 epochs and trained on target data for 500 epochs. The AdamW was utilized for optimization. The initial learning rate was set to 0.001 and decayed in a polynomial scheduler with a power of 0.9. The batch size was 2. Raw volumes were z-score normalized and scaled to the patches with the dimension of $96\times96\times96$. The performance was evaluated using Dice Similarity Coefficient (DSC;$\%$) and 95$\%$ Hausdorff distance (95HD;mm).

\begin{table*}[!t]
\centering
\caption{Performance comparison in open-set medical image segmentation. The results were reported as Mean$\pm$SD. \textbf{Bold} represents the best results. ($^*$: $p<0.01$ with the Mann-Whitney U test between ASFOSDA and other SFDA and ASFDA methods).}
\label{tab1}
\resizebox{\textwidth}{!}{
\begin{tabular}{c|c|c|cc|cc|cc}
\hline
\multirow{3}{*}{Net} & \multirow{3}{*}{Task} & \multirow{3}{*}{Methods}  & \multicolumn{4}{c|}{Open-set CT Segmentation} & \multicolumn{2}{c}{Open-set MR Segmentation} \\
\cline{4-9}
& & & \multicolumn{2}{c|}{MSD Spleen $\rightarrow$ FLARE} & \multicolumn{2}{c|}{FLARE $\rightarrow$ AMOS CT} & \multicolumn{2}{c}{Abdominal MR $\rightarrow$ AMOS MR} \\
\cline{4-9}
& & & DSC(\%)$\uparrow$ & 95HD(mm)$\downarrow$ & DSC(\%)$\uparrow$ & 95HD(mm)$\downarrow$ & DSC(\%)$\uparrow$ & 95HD(mm)$\downarrow$  \\
\hline
\multirow{14}{*}{\rotatebox[origin=c]{90}{U-Net}} 
&& Source Only        &  12.36$\pm$12.84 & 62.32$\pm$20.72 & 9.65$\pm$10.70 & 68.59$\pm$25.51 & 8.35$\pm$10.23 & 70.34$\pm$30.34 \\
&& Target Supervision &  95.21$\pm$1.40  & 1.31$\pm$1.12   & 92.24$\pm$1.64 & 1.84$\pm$1.13   & 90.16$\pm$1.21 & 1.91$\pm$1.17 \\
 \cline{2-9} & \multirow{3}{*}{\rotatebox[origin=c]{90}{SFDA}}
 & DPL      \cite{chen2021source} & 40.92$\pm$15.45 & 44.45$\pm$15.21 & 38.94$\pm$17.98 & 48.36$\pm$17.76 & 35.51$\pm$18.53 & 50.34$\pm$19.41 \\
&& FSM      \cite{yang2022source} & 43.36$\pm$15.22 & 40.07$\pm$14.23 & 40.58$\pm$16.54 & 45.42$\pm$16.91 & 36.63$\pm$16.72 & 49.27$\pm$18.71 \\
&& TT-SFUDA \cite{vs2024target}   & 47.62$\pm$14.39 & 38.31$\pm$14.98 & 43.12$\pm$15.19 & 40.18$\pm$16.55 & 40.56$\pm$16.75 & 45.64$\pm$18.38 \\
\cline{2-9}
 & \multirow{9}{*}{\rotatebox[origin=c]{90}{ASFDA}}
&  RAND                               & 68.18$\pm$10.90 & 10.94$\pm$12.11 & 66.96$\pm$12.46 & 11.97$\pm$12.25 & 61.24$\pm$13.43 & 13.18$\pm$15.63 \\
&& PENT     \cite{wang2014new}        & 72.24$\pm$10.65 & 8.24$\pm$9.26   & 71.81$\pm$10.31 & 8.67$\pm$9.36   & 65.83$\pm$11.75 & 11.92$\pm$13.56 \\
&& LC       \cite{li2006confidence}   & 71.98$\pm$10.72 & 8.45$\pm$9.38   & 71.64$\pm$11.10 & 8.69$\pm$9.83   & 65.73$\pm$11.97 & 12.17$\pm$14.15 \\
&& PMAR     \cite{wang2014new}        & 72.31$\pm$10.68 & 8.20$\pm$9.81   & 71.20$\pm$11.78 & 8.87$\pm$10.11  & 65.35$\pm$12.27 & 12.35$\pm$14.17 \\
&& Core-set \cite{sener2018active}    & 70.86$\pm$10.48 & 9.01$\pm$9.24   & 70.28$\pm$11.99 & 9.28$\pm$10.37  & 65.25$\pm$12.84 & 12.46$\pm$14.31 \\
&& MAXREP   \cite{yang2017suggestive} & 73.88$\pm$9.32  & 7.26$\pm$8.56   & 73.32$\pm$10.46 & 7.56$\pm$9.86   & 67.78$\pm$10.45 & 10.86$\pm$11.83 \\
&& CENT     \cite{liu2023cluster}     & 74.56$\pm$9.11  & 6.80$\pm$7.91   & 73.48$\pm$9.65  & 7.48$\pm$8.82   & 68.16$\pm$10.32 & 10.52$\pm$10.34 \\
&& UGTST    \cite{luo2024uncertainty} & 74.76$\pm$8.66  & 6.52$\pm$7.15   & 73.93$\pm$9.72  & 7.04$\pm$7.51   & 68.62$\pm$10.01 & 10.16$\pm$9.99   \\
&& ASFOSDA (Ours)  &  \textbf{80.11}$^*\pm$8.01 & \textbf{5.20}$^*\pm$6.22 & \textbf{78.69}$^*\pm$8.87 & \textbf{5.75}$^*\pm$6.38 & \textbf{71.46}$^*\pm$8.42 & \textbf{8.78}$^*\pm$9.82   \\
\hline
\multirow{14}{*}{\rotatebox[origin=c]{90}{Swin UNETR}} 
&& Source Only        &  11.93$\pm$13.04 & 64.84$\pm$21.77 & 9.30$\pm$10.91 & 69.00$\pm$26.24 & 8.16$\pm$10.57 & 72.31$\pm$32.42 \\
&& Target Supervision &  94.88$\pm$1.56  & 1.51$\pm$1.35   & 91.88$\pm$1.78 & 1.89$\pm$1.21   & 90.05$\pm$1.37 & 1.93$\pm$1.22 \\
\cline{2-9} & \multirow{3}{*}{\rotatebox[origin=c]{90}{SFDA}}
& DPL       \cite{chen2021source} & 39.93$\pm$16.91 & 46.56$\pm$16.76 & 37.69$\pm$18.66 & 49.33$\pm$17.89 & 33.83$\pm$18.88 & 55.34$\pm$21.34 \\
&& FSM      \cite{yang2022source} & 42.54$\pm$16.44 & 41.53$\pm$15.70 & 39.18$\pm$17.42 & 47.62$\pm$16.87 & 34.12$\pm$18.35 & 53.62$\pm$20.66 \\
&& TT-SFUDA \cite{vs2024target}   & 47.37$\pm$15.51 & 40.82$\pm$14.32 & 42.81$\pm$17.28 & 41.78$\pm$16.61 & 38.35$\pm$17.62 & 48.55$\pm$20.45 \\
\cline{2-9}
& \multirow{9}{*}{\rotatebox[origin=c]{90}{ASFDA}}
& RAND                                & 67.91$\pm$11.81 & 11.85$\pm$12.45 & 65.87$\pm$13.11 & 12.05$\pm$12.36 & 60.84$\pm$14.21 & 15.34$\pm$16.46 \\
&& PENT     \cite{wang2014new}        & 71.52$\pm$11.33 & 8.77$\pm$10.25  & 70.85$\pm$11.22 & 9.03$\pm$9.92   & 65.29$\pm$12.55 & 12.38$\pm$14.02 \\
&& LC       \cite{li2006confidence}   & 71.44$\pm$11.28 & 8.82$\pm$10.90  & 70.78$\pm$12.09 & 9.12$\pm$9.98   & 64.78$\pm$12.78 & 13.26$\pm$14.15 \\
&& PMAR     \cite{wang2014new}        & 71.85$\pm$11.12 & 8.71$\pm$10.97  & 70.17$\pm$12.91 & 9.48$\pm$10.32  & 64.52$\pm$13.32 & 13.76$\pm$15.81 \\
&& Core-set \cite{sener2018active}    & 70.59$\pm$10.68 & 9.25$\pm$9.70   & 69.99$\pm$12.63 & 10.31$\pm$10.88 & 64.24$\pm$13.71 & 13.82$\pm$15.85 \\
&& MAXREP   \cite{yang2017suggestive} & 72.62$\pm$10.51 & 7.82$\pm$8.69   & 72.34$\pm$10.58 & 8.32$\pm$9.72   & 66.62$\pm$12.23 & 11.65$\pm$12.89 \\
&& CENT     \cite{liu2023cluster}     & 73.85$\pm$10.05 & 7.44$\pm$8.72   & 72.68$\pm$10.16 & 8.23$\pm$9.60   & 66.94$\pm$11.44 & 11.54$\pm$10.82 \\
&& UGTST    \cite{luo2024uncertainty} & 73.99$\pm$9.98  & 6.88$\pm$7.94   & 72.92$\pm$9.98  & 7.88$\pm$8.15   & 67.73$\pm$11.12 & 10.93$\pm$10.32 \\
&& ASFOSDA (Ours) & \textbf{79.81}$^*\pm$9.24  & \textbf{5.64}$^*\pm$6.16   & \textbf{77.95}$^*\pm$8.95  & \textbf{5.97}$^*\pm$6.57   & \textbf{70.35}$^*\pm$9.46  & \textbf{9.32}$^*\pm$9.73 \\
\hline
\end{tabular}}
\end{table*}

\begin{figure*}[!t]
    \centering
    \includegraphics[width=\linewidth]{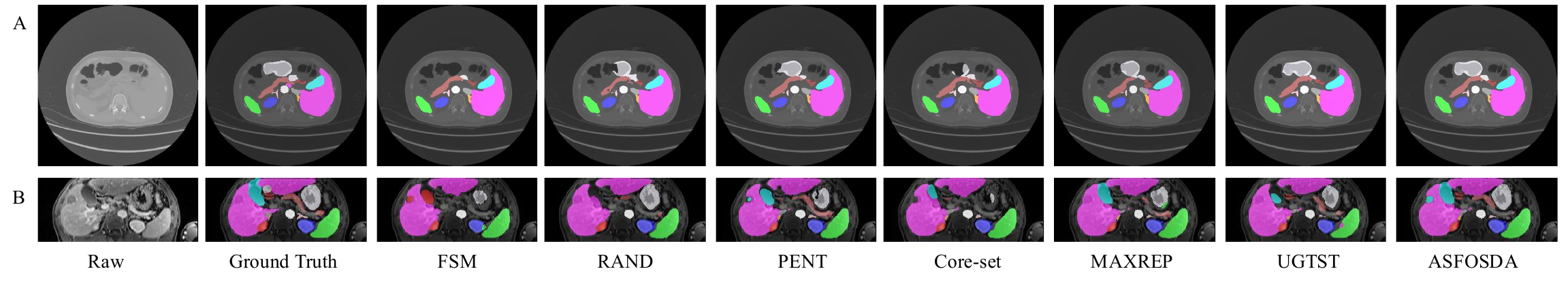} \\
    \caption{Qualitative comparison among results of SFDA, ASFDA, and our methods by adapting U-Net to (A) AMOS CT and (B) AMOS MR datasets.}
    \label{fig2}
\end{figure*}

\subsubsection{Experimental results.} We compared the performance of ASFOSDA with (i) Source Only: lower bound performance where source-trained models were employed to target domains without adaptation; (ii) Target Supervision: upper bound performance where source-trained models were trained by all target samples with annotation; (iii) SFDA methods: DPL \cite{chen2021source}, FSM \cite{yang2022source}, and TT-SFUDA \cite{vs2024target}; (iv) ASFDA methods with $5\%$ target samples queried: Random Selection (RAND), Predictive Entropy (PENT) \cite{wang2014new}, Least Confidence (LC) \cite{li2006confidence}, Probability Margin (PMAR) \cite{wang2014new}, Core-set \cite{sener2018active}, Maximum Representation (MAXREP) \cite{yang2017suggestive}, Cluster Entropy (CENT) \cite{liu2023cluster}, and UGTST \cite{luo2024uncertainty}. For better adaptation, one target sample with annotation was selected randomly for source initialization in all methods.

With a $5\%$ target sample annotation budget, our method consistently outperformed existing SFDA and ASFDA methods across both 3D U-Net and Swin UNETR (Table \ref{tab1} and Fig.~\ref{fig2}). In the MSD Spleen-to-FLARE (single to four organs) transition, our approach achieved over $84\%$ of the supervised upper bound for both networks. Similarly, for the FLARE-to-AMOS CT (four to 14 organ) task, ASFOSDA enabled the models to reach $85.31\%$ and $84.84\%$ of their respective upper bounds. Finally, in MR image adaptation (eight to 13 organs), ASFOSDA enabled U-Net and Swin UNETR to maintain robust performance, achieving $82.62\%$ and $78.12\%$ of the upper bound.

\begin{figure*}[!t]
    \centering
    \includegraphics[width=\linewidth]{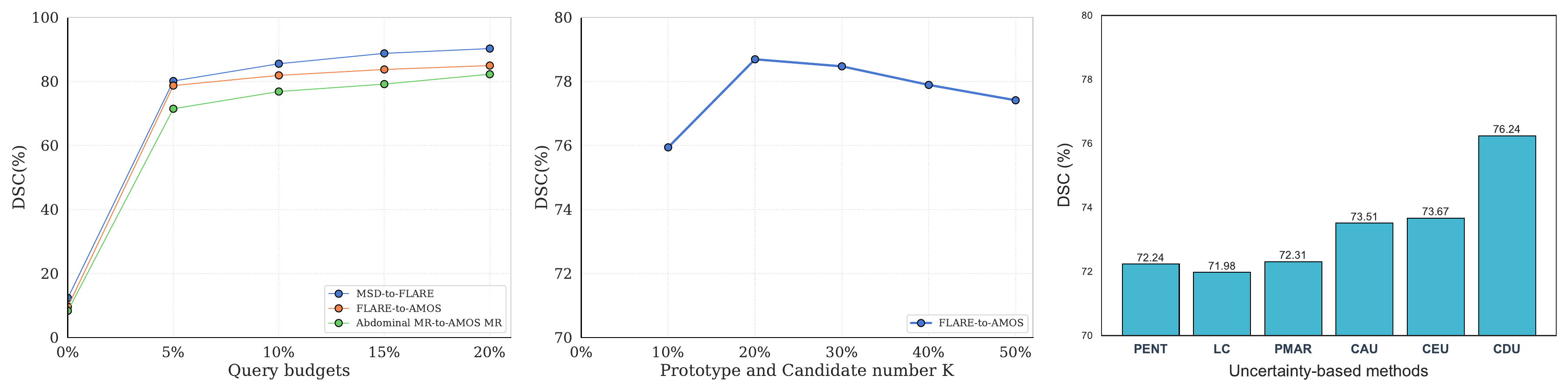} \\
    \caption{Ablation studies of key hyper-parameters and components. (A) Impact of different target budgets on U-Net adaptation. (B) Impact of different number of prototypes and candidates $K$ using $5\%$ budget for U-Net FLARE-to-AMOS adaptation. (C) Performance comparison of uncertainty methods in U-Net MSD-to-FLARE adaptation.}
    \label{fig3}
\end{figure*}

\begin{table*}[t!]
\centering
\caption{The results of ablation study on the active open-set query strategy for U-Net adaptation with $5\%$ samples queried. \textbf{Bold} represents the best results. ($^*$: $p<0.01$ with the Mann-Whitney U test between different components).}
\label{tab2}
\resizebox{\textwidth}{!}{
\begin{tabular}{cccc|cc|cc|cc}
\hline
\multicolumn{4}{c|}{Components} & \multicolumn{2}{c|}{MSD Spleen $\rightarrow$ FLARE} & \multicolumn{2}{c|}{FLARE $\rightarrow$ AMOS CT} & \multicolumn{2}{c}{Abdominal MR $\rightarrow$ AMOS MR} \\
\hline
CAU & CEU & CPD & Self-training & DSC(\%)$\uparrow$ & 95HD(mm)$\downarrow$ & DSC(\%)$\uparrow$ & 95HD(mm)$\downarrow$ & DSC(\%)$\uparrow$ & 95HD(mm)$\downarrow$\\
\hline
\checkmark &            &            &            & 73.51$\pm$9.73 & 7.42$\pm$8.61 & 72.71$\pm$10.86 & 8.17$\pm$9.34 & 66.20$\pm$11.27 & 11.81$\pm$12.44 \\
           & \checkmark &            &            & 73.67$\pm$9.58 & 7.34$\pm$8.24 & 72.82$\pm$10.72 & 8.08$\pm$9.26 & 66.32$\pm$11.41 & 11.73$\pm$12.51 \\
\checkmark & \checkmark &            &            & 76.24$\pm$8.69 & 6.06$\pm$7.12 & 74.30$\pm$9.11  & 6.85$\pm$7.40 & 69.28$\pm$9.89  & 9.66$\pm$10.02 \\
\checkmark & \checkmark & \checkmark &            & 78.46$\pm$8.34 & 5.83$\pm$6.68 & 76.48$\pm$8.82  & 6.45$\pm$7.12 & 70.55$\pm$9.44  & 9.02$\pm$9.91 \\
\checkmark & \checkmark & \checkmark & \checkmark & $\boldsymbol{80.11}^*\pm$8.01 & $\boldsymbol{5.20}^*\pm$6.22 & \textbf{78.69}$^*\pm$8.87  & \textbf{5.75}$^*\pm$6.38 & \textbf{71.46}$^*\pm$8.42 & \textbf{8.78}$^*\pm$9.82\\
\hline
\end{tabular}}
\end{table*}

\subsubsection{Ablation study.} Comparing performance of different components from Active Open-set Query strategy in three adaptation tasks demonstrated its effectiveness (Table \ref{tab2}). The performance of U-Net demonstrated consistent improvement as the query budget increased from $0\%$ to $20\%$ (Fig.~\ref{fig3}(A)). Employing $20\%$ prototypes and candidates $K$ achieved the best trade-off between segmentation accuracy and computational overhead (Fig.~\ref{fig3}(B)). Employing CAU, CEU, and CDU to evaluate uncertainty for querying target samples achieved superior performance than other uncertainty-based methods (Fig.~\ref{fig3}(C)).

\section{Conclusion}
We proposed an Active Source-Free Open-Set Domain Adaptation method for medical image segmentation. Our method achieved superior performance on volumetric multi-organ segmentation than SOTA domain adaptation methods.

\subsubsection{Acknowledgments.} The authors have no acknowledgments to declare.

\subsubsection{Disclosure of Interests.} The authors have no competing interests  to declare that are relevant to the content of this article.

%
%
%
\bibliographystyle{splncs04}
\bibliography{Paper-2507}

\end{document}